\documentclass[aps,pra,superscriptaddress,floatfix,10pt,twocolumn]{revtex4-1}
\usepackage{newtxtext}
\usepackage{graphicx,graphics}
\usepackage{dcolumn}
\usepackage{amsmath,amssymb,amsfonts}
\usepackage{latexsym,verbatim}
\usepackage{bm,dsfont}
\usepackage[caption=false]{subfig}
\usepackage{color}
\usepackage{ulem}
\usepackage{multirow}
\usepackage{overpic}
\usepackage[breaklinks=true,colorlinks,citecolor=blue,linkcolor=blue,urlcolor=blue]{hyperref}
\usepackage{array}
\usepackage{nicefrac}
\usepackage{placeins}
\usepackage{float}

\begin{document}
\title{Conveyor-belt superconducting quantum computer}

\author{Francesco Cioni}
\email{francesco.cioni@sns.it}
\altaffiliation{These authors contributed equally.}
\affiliation{NEST, Scuola Normale Superiore, I-56127 Pisa, Italy}
\author{Roberto Menta}
\email{rmenta@planckian.co}
\altaffiliation{These authors contributed equally.}
\affiliation{Planckian srl, I-56127 Pisa, Italy}
\affiliation{NEST, Scuola Normale Superiore, I-56127 Pisa, Italy}
\author{Riccardo Aiudi}
\affiliation{Planckian srl, I-56127 Pisa, Italy}
\author{Marco Polini}
\affiliation{Planckian srl, I-56127 Pisa, Italy}
\affiliation{Dipartimento di Fisica dell’Universit\`{a} di Pisa, Largo Bruno Pontecorvo 3, I-56127 Pisa, Italy}
\author{Vittorio Giovannetti}
\affiliation{Planckian srl, I-56127 Pisa, Italy}
\affiliation{NEST, Scuola Normale Superiore, I-56127 Pisa, Italy}

\begin{abstract}
The processing unit of a solid-state quantum computer consists in an array of coupled qubits, each {\it locally} driven with on-chip microwave lines that route control signals to the qubits in order to perform logical operations. This approach to quantum computing comes with two major problems. On the one hand, it greatly hampers scalability towards fault-tolerant quantum computers, which are estimated to need a number of qubits -- and, therefore driving lines -- on the order of $10^6$. On the other hand, these lines are a source of electromagnetic noise, exacerbating frequency crowding and crosstalk,  while also contributing to power dissipation inside the dilution fridge. We here tackle these two challenges by presenting a novel quantum processing unit (QPU) for a universal quantum computer which is {\it globally} (rather than {\it locally}) driven. Our QPU relies on a closed loop of superconducting qubits with always-on ZZ interactions which we dub ``conveyor belt''. Notably, this architecture requires only $\mathcal{O}(N)$ physical qubits to run a  computation on $N$ computational qubits, in contrast to previous $\mathcal{O}(N^2)$ proposals for global quantum computation. Universality is achieved via the implementation of single-qubit gates and a {\it one-shot} Toffoli gate. The ability to perform multi-qubit operations in a single step could vastly improve the fidelity and execution time of many algorithms. 
\end{abstract}

\maketitle

\section{Introduction}
Scalability in Quantum Computing (QC) refers to the ability to increase the number of qubits and other resources in a quantum system while maintaining performance, error rates, and coherence times~\cite{Mohseni2024, google2025, Fowler2012}. Achieving scalability is essential for building quantum computers capable of solving complex problems beyond the reach of classical computers.
For devices based on superconducting circuits~\cite{Gambetta2017,Martinis2019quantum,SQ2020, supremacy_PRL_2021, blais_review_RMP,bravyi2022future,ezratty2023perspective} (and other solid-state platforms), scalability is fundamentally constrained, among others, by the need for localized drive of each physical qubit~\cite{You2005supercond, Girvin2008, devoret2013supercond, wedin2017quantum} in order to perform single-qubit rotations and entangling operations as required by the computation. This issue, often called the “wiring problem", leads to a wiring overload~\cite{Gambetta2017, tsai2020pseudo, tsai2021gate}: for example, if we had to scale the current microwave interface to control a million-qubit system, the electronics itself would take up to three football fields of space, while consuming about 40 MW of dc power~\cite{Bardin2020}.
A potential solution to this problem is the development of globally driven QC schemes~\cite{Lloyd_1993,benjamin_2000,benjamin_2001,benjamin_2003, benjamin_2004}. These schemes involve designing and operating quantum computers in a manner that minimizes or eliminates the need for precise, localized manipulation of individual qubits. The main benefits include the simplification of designs due to the reduced complexity of control electronics and interconnections, and potentially enhanced error resilience with respect to local errors.
However, globally driven QC approaches also face challenges, such as ensuring precision in global operations, creating and maintaining large, possibly highly entangled states for logical information encoding, and implementing fault-tolerant operations globally, which require sophisticated error-correcting protocols.

A recently proposed globally driven superconducting QC architecture~\cite{menta2024globally} addresses some of these challenges, specifically the precision in global operations and the effectiveness of state preparation.
Building upon a proposal based on Rydberg atoms~\cite{cesa2023universal, fromonteil2024hamilton}, in Ref.~\cite{menta2024globally}, we demonstrated that a globally driven universal quantum computer can be achieved on a superconducting platform by 
leveraging the longitudinal (always-on) ZZ interaction~\cite{ZZ-suppression2020, Xu2024, ZZ-contrast2020, ZZ-eliminating2022,ZZ_2020, ZZ_2021, xu2021z, Fors2024} between neighboring superconducting qubits organized in a two-dimensional (2D) ladder. In this work, we present a novel architecture that significantly improves the design proposed in Ref.~\cite{menta2024globally} by drastically reducing the resources needed to implement it. 
\begin{figure*}[t!]
\centering
\includegraphics[width=1.0\textwidth]{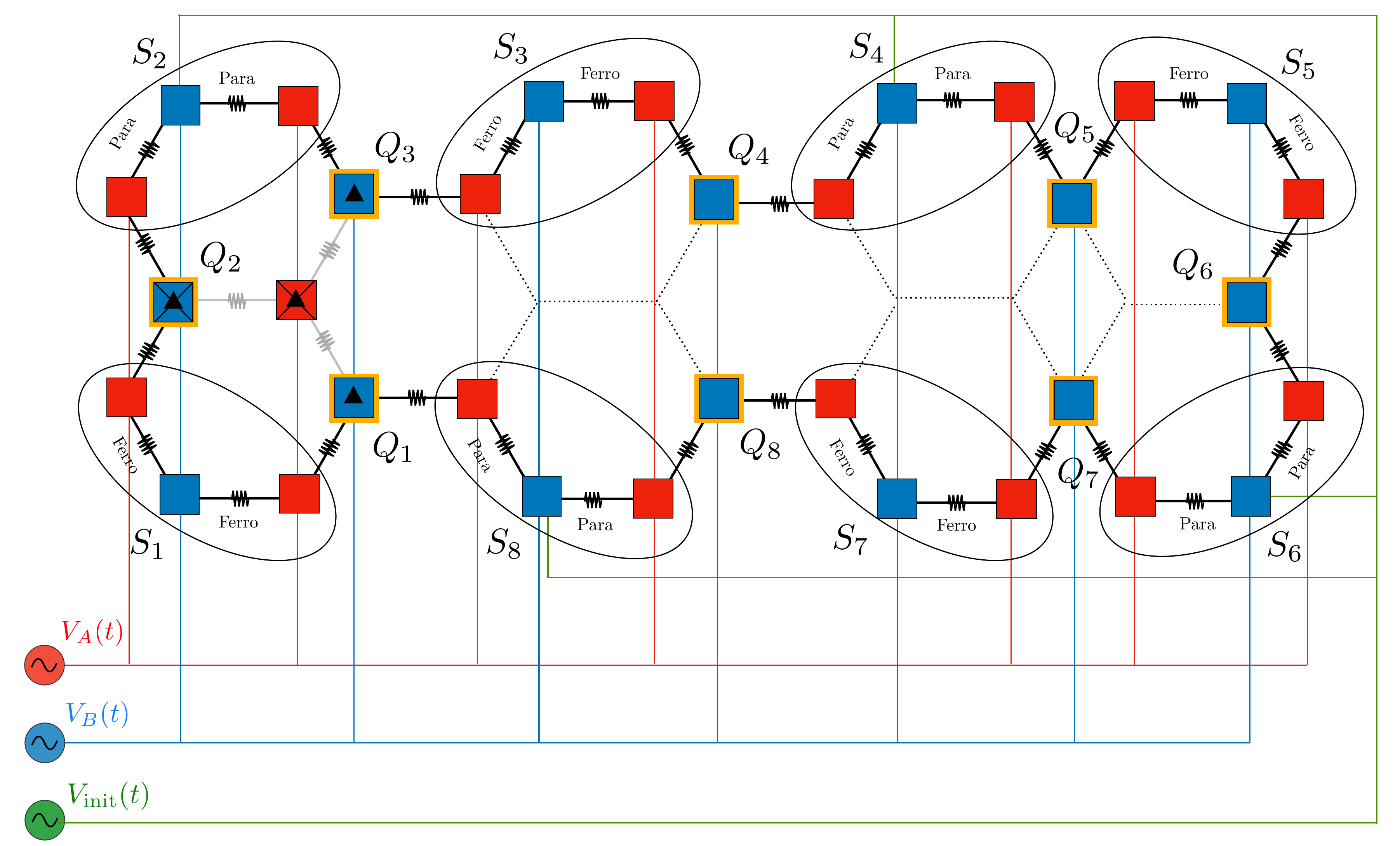}
\caption{\label{fig:architecture}Schematic description of the proposed architecture. Two types, $A$ and $B$, of superconducting qubits (red and blue squares, respectively) are separately driven by two classical sources $V_{A,B}(t)$  (red and blue continuous lines). They are coupled via a longitudinal ZZ coupling (black and grey springs). Black triangles inside regular and crossed qubits denote local corrections in the resonance frequency of the qubits (see main text). The $A$-type crossed qubit (red square inside  the loop) enables one-shot Toffoli gate (three-qubit gate) -- the corresponding 
interactions are depicted in gray. The $B$-type crossed qubit performs single-qubit gates. 
The elements highlighted in yellow indicate the information carrying sites  $Q_1$, $Q_2$, $\cdots$, $Q_N$, separated
by three-qubits sectors $S_1$, $S_2$, $\cdots$, $S_N$.
The $Q_j$'s host the computational qubits through well-formed configurations (\ref{encoding}) where the $S_j$'s are initialized in alternating
sequences of  paramagnetic ($geg$) and ferromagnetic ($ggg$)
phases. Since ZZ interactions critically depend on the relative distance between two qubits, we have used the honeycomb structure to make sure that the couplings among adjacent qubits are (nominally) all identical.
The figure pertains to a $N=8$ qubit quantum computer.}
\end{figure*}
At variance with previous schemes~\cite{menta2024globally, cesa2023universal}, which require to operate globally on a 2D array of ${\cal O}(N^2)$ physical qubits to run universal operations on $N$ computational qubits, 
the setup discussed here only needs ${\cal O}(N)$ elements -- specifically $4N+1$ -- organized in a closed loop, as shown in Fig.~\ref{fig:architecture}. For clarity, we stress that the computational qubits are the ones in which we encode the logical information. On the other hand, the set of physical qubits encompasses all two-level systems present in the architecture.
Sequences of control pulses that act collectively on half of the elements of the device allow us to induce special permutations on the states of the sites. Combining these transformations with the possibility to implement both single-qubit gates and a one-shot Toffoli gate yields universal quantum computation. It is important to mention that the associated dynamical processes are based on a blockade regime, analogous to the one in Rydberg atoms~\cite{Rydberg_blockade}, which is here achieved thanks to a strong ZZ coupling between nearest-neighbor qubits. The main features of this effect for the case of superconducting qubits are described in Ref.~\cite{menta2024globally}. We also mention that, in QC, the ability to perform multi-qubit operations, such as the Toffoli gate, in a {\it single} step could vastly improve the fidelity and execution time of many algorithms~\cite{Zahedinejad2015, baker2022}.
To the best of our knowledge, our proposal introduces a novel superconducting platform that, for the first time, incorporates a {\it one-shot} Toffoli gate~\cite{Nielsen2010}, which operates based on the ZZ-induced blockade regime.

\section{Model}
The system shown in Fig.~\ref{fig:architecture} consists of a closed loop formed by $4N$ superconducting qubits coupled via a spatially-uniform ZZ interaction of  coupling strength $\hbar \zeta$ (black springs  in the figure). The loop contains $N$ sites, identified by the symbols $Q_1$, $Q_2$, $\cdots$, $Q_N$, which play a special role in the model and will be referred to as {\it Information Carrying} (IC) sites. For all $j\in \{1,\cdots,N\}$ the IC site $Q_j$ is separated from its adjacent counterpart $Q_{j+1}$ by a sector $S_j$ which contains three “non-active" sites.
In the figure, there is an extra superconducting qubit placed inside the loop, which is connected to the first three IC sites $Q_1$, $Q_2$, and $Q_3$ via the same ZZ coupling term that connects every other element (these three couplings are highlighted in a light gray color to emphasize the fact that this additional qubit is placed inside the loop and not along it). The superconducting qubits in the model belong to two distinct families: $A$-type qubits, represented as red squares, and $B$-type qubits, represented as blue squares, which alternate in the loop in an $ABA$ pattern. As shown in the figure, all the IC sites host a $B$-type qubit, while the intermediate sectors $S_j$ are formed by one $B$-type and two $A$-type qubits. All qubits of the same type share the same level spacing $\hbar \omega_{A/B}$, except for the elements marked with triangles, which, for reasons that will be clear in the following, require $\zeta$-dependent local corrections to their level spacings i.e.~$\hbar \omega_{A/B} \to \hbar (\omega_{A/B} + \zeta)$, due to the higher number of nearest neighbors they have (three) compared to the other qubits in the setup. To eliminate the possibility of a residual “swap" term between nearest-neighbor qubits, it may be necessary to have $\omega_A \neq \omega_B$.
Each family of qubits is collectively driven by the same time-dependent external source, termed $V_{A/B}(t)$. Thus, the control is global. The device also includes two inhomogeneities: the additional $A$-type element previously mentioned and one of the three $B$-type elements connected to it (specifically, the one located at the IC site $Q_2$). These special qubits, referred to as \textit{crossed qubits}, are used to perform multi- and single-qubit operations. They maintain the nominal level spacing of their respective families but have an augmented coupling with the external source, i.e. the Rabi frequency. As proved in Ref.~\cite{menta2024globally}  and summarized in Section \ref{sec:GCD} and more in detail in Appendix \ref{app:global_control}, this difference allows independent control of the normal qubits and crossed qubits of a given family, despite the global control pulse acting simultaneously on them. Both the $A$- and $B$-type crossed qubits are marked with a small black triangle.
Finally, a third control line, $V_{\mathrm{init}}(t)$, acting on a subset of the $B$-type qubits, is necessary for the initialization of the machine. This control line is active only at the initial stage of the computation. Regarding the readout procedure, a possible implementation is discussed in Appendix~\ref{sec:init-readout}.

Adopting the same convention of Ref.~\cite{menta2024globally} we write the Hamiltonian of the setup as $\hat{H}(t): = \hat{H}_0 + \hat{H}_{\rm drive}(t)$, where
\begin{equation} 
\hat{H}_0 := \sum_{\chi \in \{A,B\}} \sum_{i \in \chi} \frac{\hbar \omega_i}{2} \hat{\sigma}_i^{(z)}
+  \sum_{\langle i,j \rangle} \frac{\hbar \zeta}{2} \hat{\sigma}_i^{(z)} \otimes \hat{\sigma}_j^{(z)}
\;, 
\end{equation} 
describes the local energy contribution of the superconducting qubits and their ZZ interactions which are fixed by the geometry of the model, while 
\begin{equation} 
\hat{H}_{\rm drive}(t) := \sum_{\chi \in \{A,B\}} \sum_{i \in \chi} \hbar \Omega_\chi(t)\sin(\omega_{\mathrm{d},\chi}t + \phi_{\chi}(t)) \hat{\sigma}_i^{(y)}
\;,
\end{equation} 
is the time-dependent driving contribution induced by the classical control lines.
In these equations, $\hat{\sigma}_i^{(x,y,z)}$ represent ordinary Pauli matrices acting on the Hilbert space of the $i$-th qubit, expressed in the local energy basis $\{\vert g_i \rangle, \vert e_i \rangle\}$. The summation in the interacting part of 
$\hat{H}_{0}$ encompasses all nearest-neighbor interactions, mathematically representing all the black springs in Fig.~\ref{fig:architecture}. The parameter $\omega_{\mathrm{d},\chi}$ denotes the oscillation frequency of the driving pulse $V_\chi(t)$, while $\Omega_\chi(t)$ and $\phi_\chi(t)$ define the time-dependent Rabi frequency and phase of such control. In the following, we will assume that these quantities assume 
constant values on disjoint time windows, in such a way that, at each time, only one species (either $A$ or $B$) is driven. 
For simplicity, in writing $\hat{H}_{\rm drive}(t)$, we have omitted the driving term associated with the control line $V_{\rm init}(t)$, which operates solely at the very beginning of the computational process on a specific subset of the physical qubits. Additionally, we did not explicitly highlight that when the site index $i$ in the expression for  $\hat{H}_0$ identifies a $B$-type qubit marked with a black triangle, the corresponding qubit's level spacing becomes $\hbar(\omega_B+\zeta)$ instead of $\hbar\omega_B$. The same substitution must be performed when the site index $i$ identifies an $A$-type qubit marked with a black triangle, i.e.~an $A$-type crossed qubit. Similarly, whenever in 
the expression of $\hat{H}_{\rm drive}(t)$ 
the index $i$ identifies a crossed qubit, $\Omega_{\chi}(t)$ needs to be replaced by $2\Omega_\chi(t)$.
Apart from these local adjustments (which need to be engineered once and for all at fabrication level), it is important to note that $\Omega_\chi(t)$, $\phi_\chi(t)$, and $\omega_{\mathrm{d},\chi}$ are independent of the site index $i$, indicating that they are associated with a control pulse acting globally on all qubits of $\chi$-type in the model.
We conclude this paragraph by stressing that the above Hamiltonian is exactly the same as the one studied in Ref.~\cite{menta2024globally}. However, due to the more exotic geometry, the dynamical features of the system are more complex than in the case of the 2D ladder geometry used in Ref.~\cite{menta2024globally}. We take advantage of this complexity to achieve the $\mathcal{O}(N)$ scaling, as opposed to the $\mathcal{O}(N^2)$ scaling of Refs.~\cite{menta2024globally,cesa2023universal}.

\section{Dynamical features and global control}
\label{sec:GCD}

The present proposal relies on two key ingredients.
The first is the emulation of a Rydberg-blockade-like mechanism~\cite{Rydberg_blockade} through the $ZZ$ interactions between qubits.  
The second is the presence of two distinct Rabi frequencies within each qubit family, corresponding to regular and crossed qubits, which enables independent control despite being driven by the same global control.

\subsection{Effective blockade dynamics}

We consider global pulses $(\Omega_\chi(t), \phi_\chi (t))$ with $\chi=\{A,B\}$, applied piecewise in time such that only one qubit species is driven at any given instant.  
One can show that under the strong-coupling
  condition $\eta_{\rm BR}:=|\zeta/ \Omega_{\chi}|\gg 1$, by properly detuning  the driving frequencies 
  $\omega_{\mathrm{d},\chi}$
 from the nominal level spacing of the qubits, one is able to selectively induce transitions among states of the system
only when two nearest-neighbour sites do not simultaneously occupy their excited levels. 
As a result, the dynamics is effectively restricted to configurations in which no two interacting neighbors occupy the excited state, realizing a blockade mechanism analogous to that of Rydberg systems. 

In this limit, the unitary evolution $\hat{U}_\chi$ generated by a single global pulse of length $\tau$ acting on qubits of type $\chi$ can be written as~\cite{menta2024globally}
\begin{equation}
\hat{U}_\chi
=
\prod_{i\in\chi}
\left[
\hat{\openone}_i\otimes\hat{Q}_{\langle i\rangle}
+
\hat{\mathbb{R}}_i(\theta, \bm{n}_\perp)\otimes\hat{P}_{\langle i\rangle}
\right],
\label{eq:Uchi_blockade}
\end{equation}
up to an overall phase~\cite{projectors}.
Here $\hat{P}_{\langle i\rangle}$ projects onto the subspace in which all nearest neighbors of site $i$ are in the ground state, $\hat{Q}_{\langle i\rangle}=\openone-\hat{P}_{\langle i\rangle}$, and
\begin{equation}
\hat{\mathbb{R}}_i(\theta, \bm{n}_\perp)
=
\exp\!\left[
-i\frac{\theta}{2}
\bm{n}_\perp\!\cdot\!\vec{\sigma}^{(i)}
\right]
\end{equation}
is the single-qubit rotation induced in the absence of interactions, with angle
$\theta=\Omega_\chi\tau$ and axis
$\bm{n}_\perp=(\cos\phi_\chi,\sin\phi_\chi,0)$.

\subsection{Global control unitaries}

Due to the geometry of the device, qubits of the same species $\chi$ split into two subsets defined as regular ($\chi^{\rm r}$) and crossed ($\chi^{\times}$). As mentioned, these qubits are characterized by different Rabi frequencies.  
It can be shown that this inhomogeneity is sufficient to achieve independent control of the two subsets, even though the pulses acting on the qubits are global~\cite{menta2024globally}.

Due to the global nature of the drive, the evolution induced by a single pulse can be written compactly as
\begin{equation}\label{eq:elem_pulse}
\hat{U}_\chi
=
\hat{W}_\chi(\theta,\bm{n}_\perp;2\theta,\bm{n}_\perp),
\end{equation}
where we define the global control unitaries
\begin{equation}
\hat{W}_{\chi}(\theta',\bm{n}';\theta'',\bm{n}'')
=
\hat{W}_{\chi^{\rm r}}(\theta',\bm{n}')
\hat{W}_{\chi^{\times}}(\theta'',\bm{n}'').
\label{eq:ccugen}
\end{equation}
Recalling the above mentioned result, for $\xi\in\{\chi^{\rm r},\chi^{\times}\}$, we have
\begin{equation}
\hat{W}_{\xi}(\theta,\bm{n})
=
\prod_{i\in\xi}
\left[
\hat{\openone}_i\otimes\hat{Q}_{\langle i\rangle}
+
\hat{\mathbb{R}}_i(\theta,\bm{n})\otimes\hat{P}_{\langle i\rangle}
\right],
\end{equation}
with
$\hat{\mathbb{R}}_i(\theta,\bm{n})
=\exp[-i(\theta/2)\bm{n}\cdot\vec{\sigma}_i]$
a single-qubit $SU(2)$ rotation~\cite{Nielsen2010}.
Depending on the state of the neighboring sites, each qubit either undergoes the rotation $\hat{\mathbb{R}}_i(\theta,\bm{n})$ or remains unchanged.

When only one of the parameters $\theta'$ or $\theta''$ in Eq.~\eqref{eq:ccugen} is nonzero, the corresponding transformation acts selectively on either the regular or crossed subset of qubits.  
Although the elementary pulses in Eq.~\eqref{eq:elem_pulse} produce correlated rotations on the two subsets, suitable pulse sequences allow one to synthesize arbitrary rotations on $\chi^{\rm r}$ and $\chi^{\times}$ independently, and hence to generate generic unitaries of the form $\hat{W}_{\chi}(\theta',\bm{n}';\theta'',\bm{n}'')$.  
The explicit constructions establishing the universality of such control are summarized in Appendix~\ref{app:global_control} and detailed in Ref.~\cite{menta2024globally}. A generalization of this result to the case of more than two subsets is proven in Ref.~\cite{menta2025building}.

\section{Information encoding and exchange operations}

\begin{figure*}[t]
\centering
\includegraphics[width=1.0\textwidth]{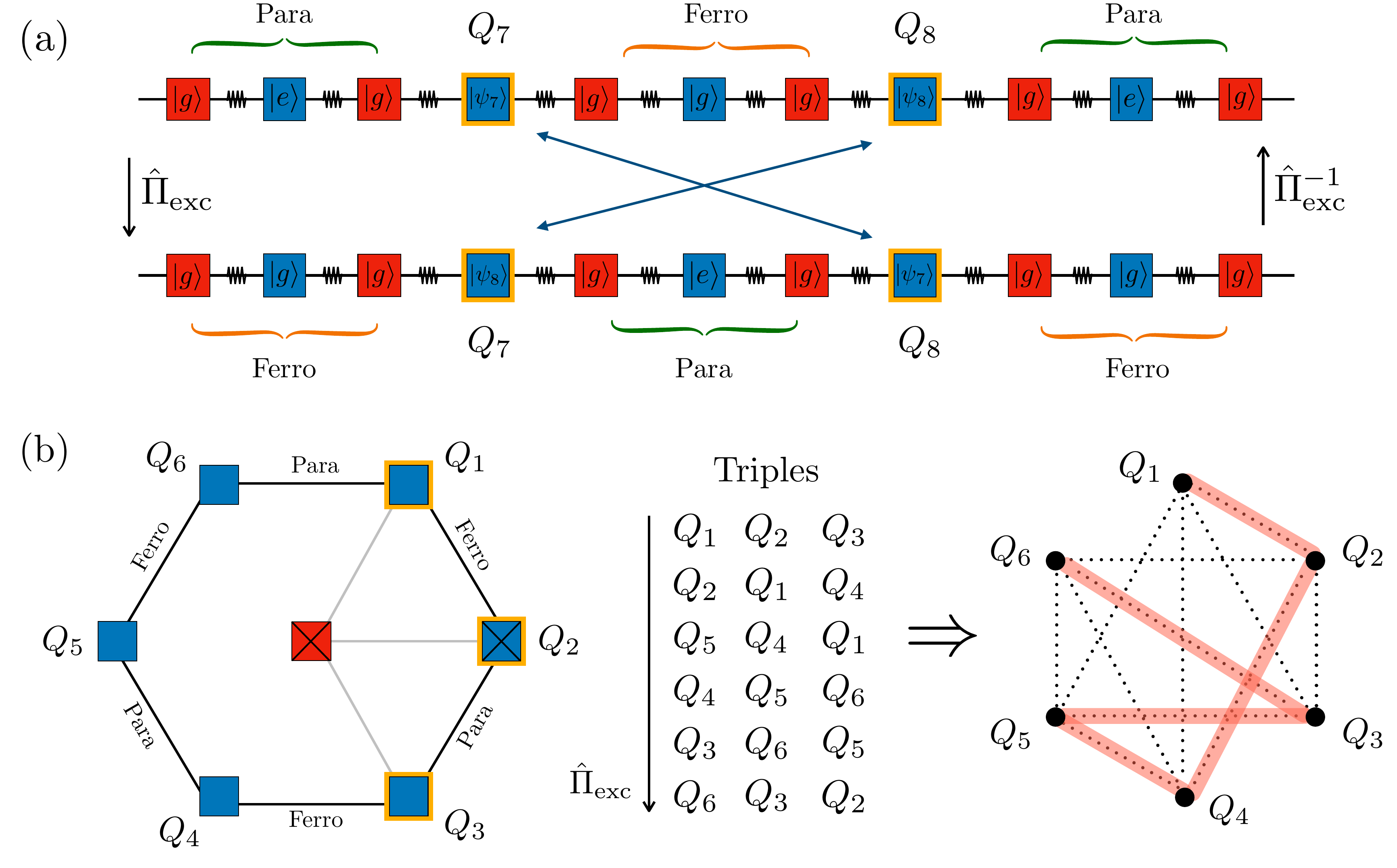}
\caption{\label{fig:swap} (a) Schematic description of the  unitary evolution induced by $\hat{\Pi}_{\rm exc}$ on a ${\rm FP}$ well-formed state 
where the $Q_7$ and $Q_8$ sites are in the input states $\vert \psi_{7 ,8} \rangle$. This unitary operation acts as a swap between two qubits, separated by a ferromagnetic region. Moreover, after the unitary, paramagnetic and ferromagnetic phases are exchanged. The reverse unitary transformation is $\hat{\Pi}^{-1}_{\rm exc} := \hat{\Pi}_B \hat{\Pi}_{\rm exc} \hat{\Pi}_B$. (b) Schematic representation of the universality proof.} 
\end{figure*}

In the setup of Fig.~\ref{fig:architecture}  the logical information is encoded in the IC sites $Q_1$,  $Q_2$, $\cdots$, $Q_N$ with the intermediate sectors  $S_1$, $S_2$, $\cdots$, $S_N$ acting as separators.  This is a peculiarity of the present quantum computing processor, which has no analogue
in the proposals of Refs.~\cite{menta2024globally,cesa2023universal}, where the string of the $N$ computational qubits can instead rigidly drift along
the entire 2D array of physical qubits of the device.
Specifically, a generic  $N$-qubit logical state $|\Psi\rangle= \sum_{\vec{k}\in \{ e,g\}^N}  \Psi_{\vec{k}} |k_1,k_2,\cdots, k_N\rangle$ is  expressed in one of the
two possible  {\it well-formed}  configurations,  $|\Psi;{\rm FP}\rangle$ or $ |\Psi;{\rm PF}\rangle$. 
Both these vectors have the central crossed $A$-type qubit in the ground state, while 
 the intermediate sectors $S_j$ are in an alternating sequence 
 of
 “ferromagnetic" ($|{\rm F}\rangle := \vert g g g \rangle$) or  “paramagnetic" ($|{\rm P}\rangle := \vert g e g \rangle$) phases.
 In particular, $|\Psi;{\rm FP}\rangle$  (resp. $|\Psi;{\rm PF}\rangle$)
initializes the sector $S_j$ in the state $|{\rm F}\rangle_{S_j}$ if $j$ is odd (even), and in   $|{\rm P}\rangle_{S_j}$ if $j$ is even (odd), so that
\begin{eqnarray}\nonumber
  &|\Psi;{\rm FP}\rangle& := \!\!\! \!\!\!\sum_{\vec{k}\in \{ g,e\}^N}  \Psi_{\vec{k}}
 |k_1\rangle_{Q_1} |{\rm F} \rangle_{S_1} |k_2\rangle_{Q_2} |{\rm P}\rangle_{S_2}
 |k_3\rangle_{Q_3} |{\rm F}\rangle_{S_3} \nonumber \\
 &\cdots& |k_{N-1}\rangle_{Q_{N-1}}  |{\rm F} \rangle_{S_{N-1}}  |k_N\rangle_{Q_N}  |{\rm P} \rangle_{S_N} \otimes |g\rangle_{A^{\times}}, \label{encoding} 
\end{eqnarray}
and $|\Psi;{\rm PF}\rangle =  |\Psi;{\rm FP}\rangle|_{|{\rm P}\rangle\leftrightarrow |{\rm F}\rangle}$. We notice that this requires $N$ to be an even number. When all qubits are initially in the ground state, the control $V_{\rm init}(t)$ can be used to initialize the system to a vector of the form $|\Psi; {\rm FP}\rangle$, see Fig.~\ref{fig:architecture}. This process brings all the associated qubits into the $|e\rangle$ state. As mentioned before, in our architecture, single and multi-qubit gates can then be performed on specific sites which host, or which are directly coupled to, crossed qubits elements (i.e.~$Q_1$, $Q_2$ and $Q_3$). This implies that 
a fundamental prerequisite to perform quantum computation in our conveyor-belt architecture is the ability to coherently exchange the positions of the computational qubits, i.e.~change the positions of the qubit states without modifying them. This is achieved trough a sequence of eight alternating global pulses, i.e. 
$\hat{\Pi}_{\rm exc} := \hat{\Pi}_B \hat{\Pi}_{A^{\rm r}} \hat{\Pi}_B \hat{\Pi}_{A^{\rm r}} \hat{\Pi}_B \hat{\Pi}_{A^{\rm r}} \hat{\Pi}_B \hat{\Pi}_{A^{\rm r}}$ with $\hat{\Pi}_{A^{\rm r}} := \hat{W}_{A^{\rm r}}(\pi, \bm{x})$ acting as a conditional-bit-flip only on the regular $A$-type qubits, and $\hat{\Pi}_{B}:= \hat{W}_{B}(\pi,\bm{x}; \pi ,\bm{x})$ acting as a conditional bit-flip on all $B$-type qubits, including the crossed one. As shown in Appendix \ref{sezDYN}, when acting on a ${\rm FP}$ (resp. ${\rm PF}$) well-formed state $|\Psi; {\rm FP}\rangle$  ($|\Psi; {\rm PF}\rangle$), the transformation $\hat{\Pi}_{\rm exc}$ runs in parallel $N$ two-qubit swap gates $\hat{U}^{{\rm swap}}_{Q_jQ_{j+1}}$ on  the pairs $\{Q_1,Q_2\}$, $\{Q_3,Q_4\}$, $\cdots$, $\{Q_{N-1},Q_{N}\}$, (resp. $\{Q_2,Q_3\}$, $\{Q_4,Q_5\}$, $\cdots$, $\{Q_{N},Q_{1}\}$) while exchanging the ferromagnetic and paramagnetic phases of the intermediate sectors producing a 
${\rm PF}$ (resp. ${\rm FP}$) output configuration---see Fig.~\ref{fig:swap}. 
The resulting motion of the computational qubits acquires an interesting feature. Indeed one has that 
$\hat{\Pi}_{\rm exc} |\Psi; {\rm FP}\rangle = |\Psi^{(1)}_{{\circlearrowright}_o,{\circlearrowleft}_e}; {\rm PF}\rangle$, and  $\hat{\Pi}_{\rm exc} |\Psi; {\rm PF}\rangle = |\Psi^{(1)}_{{\circlearrowleft}_o,{\circlearrowright}_e}; {\rm FP}\rangle$, 
with the vector $|\Psi^{(\ell)}_{{\circlearrowright}_o,{\circlearrowleft}_e}\rangle$ (resp. $|\Psi^{(\ell)}_{{\circlearrowleft}_o,{\circlearrowright}_e}\rangle$) obtained by applying to $|\Psi\rangle$ an $\ell$-step clock-wise rotation of the internal states of the sites $Q_j$ with odd  (even) index $j$ and, at the same time, an $\ell$-step anti-clock-wise rotation of the internal states of the sites $Q_j$ with even (odd) index $j$.
Notice that since the direction of the rotations depends on the initial location of the ferromagnetic and paramagnetic phases, multiple applications of $\hat{\Pi}_{\rm exc}$ pulses
do not cancel out (see Appendix \ref{sezDYN}). 
For instance, using $\ell$  times the transformation on the input state $|\Psi; {\rm FP}\rangle$ leads to 
$\hat{\Pi}^{\ell}_{\rm exc} |\Psi; {\rm FP}\rangle = |\Psi^{(\ell)}_{{\circlearrowright}_o,{\circlearrowleft}_e}; {\rm FP}\rangle$ for
$\ell$ even, and 
$\hat{\Pi}^{\ell}_{\rm exc} |\Psi; {\rm FP}\rangle = |\Psi^{(\ell)}_{{\circlearrowright}_o,{\circlearrowleft}_e}; {\rm PF}\rangle$ for $\ell$ odd (of course for $\ell=N$ the system goes back to the initial configuration).
Exploiting this feature, we can  move any IC qubit in any other IC qubit location. Suppose, now, that the aim is to apply the single qubit transformation $\hat{\mathbb{R}}(\theta,\bm{n})$ to the computational qubit  located
(say) in the $Q_j$ site of the loop. 
  Given our previous explanations, this task is straightforward. 
 We simply need to use the transformation $\hat{\Pi}^{\ell}_{\rm exc}$ with $\ell$ such that 
 the state of $Q_j$ rotates into the position $Q_2$ which is hosting the $B$-type crossed qubit. Once there, we apply the pulse $\hat{U}_{B^{\times}} := \hat{W}_{B^{\times}}(\theta, \bm{n})$ and (in case we need it), reverse the exchange operation via
 the sequence  $\hat{\Pi}^{N-\ell}_{\rm exc}$ (or using $\ell$-times the inverse of $\hat{\Pi}^{\ell}_{\rm exc}$ defined in the caption of Fig.~\ref{fig:swap}). This demonstrates that we are capable of performing any single-qubit unitary operation on any one of the computational qubits. In the next paragraph, we describe how a three-qubit Toffoli gate can be implemented, thereby achieving a universal gate set.
\section{One-shot Toffoli gate}
In common QC architectures, the Toffoli gate is performed using two-qubit gates. Here, we implement a three-qubit {\it one-shot} Toffoli gate by using a total number of four qubits: one $A$-type crossed qubit located inside the loop, which acts as a mediator, and the three $B$-type computational qubits located in the  IC sites $Q_1$, $Q_2$ and $Q_3$. A similar setup has already been proposed by Rasmussen {\it et al}.~\cite{Rasmussen202_toffoli} as an isolated multi-qubit gate. The most notable feature of our Toffoli gate is that it is naturally embedded into our conveyor-belt QC architecture.
Here, the natural way to implement a Toffoli gate is to decompose it into a controlled-controlled-Z (CCZ) gate combined with two single-qubit Hadamard gates. Since we have already discussed how to perform single-qubit gates, we simply need to explain now how to perform a CCZ gate.
The fundamental observation here is that the unitary $\hat{\mathcal{Z}}_{A^{\times}}:=\hat{W}_{A^{\times}}(2\pi, \bm{n})$  induces a $(-1)$ phase factor on the state of the system if and only if the three $B$-type qubits connected to it are all in the ground state $|g\rangle$ which, apart from  a global NOT, is exactly a CCZ gate applied on the three $B$-type qubits. Specifically, the transformation $\hat{\mathcal{Z}}_{A^{\times}}$ is obtained from~(\ref{eq:ccugen}) for $\chi=A$ and by setting $\theta'=0$ and $\theta'' = 2\pi$, independently from the choices of $\bm{n}'$ and $\bm{n}''$.
It corresponds to 
 $\hat{\mathcal{Z}}_{A^{\times}} = \hat{Q}_{\langle A^{\times} \rangle} - \hat{P}_{\langle A^{\times} \rangle}$, where, since the $A$-type crossed qubit has three connections, the projector $\hat{P}_{\langle A^{\times} \rangle}$ can be written as $\hat{P}_{\langle A^{\times} \rangle}:= |g g g\rangle\langle g g g|$, while the complementary projector is $\hat{Q}_{\langle A^{\times} \rangle}:= |eee \rangle\langle eee| +
|eeg \rangle\langle eeg| + \cdots + |gge \rangle\langle gge|$. Accordingly, setting $\bm{n}=(1,0,1)/\sqrt{2}$, the transformation
\begin{equation}\label{eq:Toffoli_explicit}
   \hat{T}_{13\rightarrow 2}:= \hat{W}_{B^{\times}}(\pi,\bm{n})  \hat{W}_{B}(\pi,\bm{x})
    \hat{\mathcal{Z}}_{A^{\times}}  \hat{W}_{B}(\pi,\bm{x}) \hat{W}_{B^{\times}}(\pi,\bm{n}) \ ,
\end{equation}
corresponds to a Toffoli gate where $Q_1$ and $Q_3$ act as controllers and $Q_2$ as the controlled qubit. In Eq.~(\ref{eq:Toffoli_explicit}), $\hat{W}_{B}(\pi,\bm{x})$ stands for $\hat{W}_{B}(\pi,\bm{x};\pi,\bm{x})$.

We conclude this paragraph by remarking that this whole discussion can be generalized to an arbitrary number of qubits connected to the central $A$-type crossed qubit. It is therefore possible, at least in principle, to realize $N$-qubit Toffoli gates in our globally-controlled conveyor-belt architectures.

\section{Universality}\label{sec:universality}
In the previous paragraphs, we have demonstrated how, by using the global controls $V_A(t)$ and $V_B(t)$, the architecture in Fig.~\ref{fig:architecture} allows us to implement any single-qubit operation on each site that encodes the logical information, as well as to perform  
the exchange gate
 $\hat{\Pi}_{\rm exc}$ and the Toffoli gate  $\hat{T}_{13\rightarrow 2}$.
To prove that this is sufficient to guarantee universal QC in the model,  we now show that, using these gates, it is possible to induce arbitrary swap gates $\hat{U}^{{\rm swap}}_{Q_jQ_{j'}}$ between any two IC sites. These will then be used to convert  $\hat{T}_{13\rightarrow 2}$ into Toffoli gates that couple all possible triples formed by computational qubits. 
It should first be noted that the Toffoli gate, when combined with single-qubit operations, can be employed to implement local swaps between any two qubits that are simultaneously connected to the central $A$-type crossed qubit~\cite{Nielsen2010}. Starting from $\hat{T}_{13\rightarrow 2}$, we can generate swap gates  between each of the pairs $\{ Q_1,Q_2\}$, $\{ Q_2,Q_3\}$, and $\{ Q_3,Q_1\}$ that we represent as edges of
a $N$ vertex graph formed by the computational qubits of the system, see  Fig.~\ref{fig:swap}(b).
Next, we apply $\hat{\Pi}_{\rm exc}$ to rotate the computational qubits, e.g.~inducing the mapping  $(Q_1,Q_2,Q_3)\leftrightarrow (Q_2,Q_1,Q_4)$, and repeat the whole procedure obtaining two new  
 swaps among  the couple $\{ Q_1,Q_4\}$ and $\{ Q_2,Q_4\}$ which allows us to draw two new edges in the graph.
 Note also that, by combining these swaps, we can also induce an extra new swap between 
 $\{ Q_3,Q_4\}$ (e.g.~$\hat{U}^{{\rm swap}}_{Q_3Q_{4}}=\hat{U}^{{\rm swap}}_{Q_1Q_{3}}\hat{U}^{{\rm swap}}_{Q_1Q_{4}} 
 \hat{U}^{{\rm swap}}_{Q_1Q_{3}}$), making the subgraph formed by the sites $Q_1$, $Q_2$, $Q_3$ and $Q_4$ fully connected.
 Proceeding along this way, one can show by recursion that all the other vertices can also be included in a fully connected graph, meaning that we can generate all possible swaps  $\hat{U}^{{\rm swap}}_{Q_jQ_{j'}}$, hence proving the thesis (see Appendix \ref{secENC} for details).

\section{Discussion}
The model we proposed represents a clear improvement in terms of scalability compared to those discussed in Refs.~\cite{menta2024globally,cesa2023universal} where  $\mathcal{O}(N^2)$ physical qubits are needed in order to have a universal quantum computer with $N$ computational qubits. Indeed, in the present work, we have presented a quantum processing unit where the scaling is $\mathcal{O}(N)$. Globally-driven schemes with linear scaling have already been proposed~ \cite{Lloyd_1993,benjamin_2000,benjamin_2001,benjamin_2003, benjamin_2004}. In contrast to our proposal, however, these ones remain rather abstract, since they do not provide precise recipes on how to actually implement the proposed global schemes. Additionally, there are many elements of difference in how the computation is performed. Among these, the processor described in the present work does not rely on the dynamical tuning of the system's parameters, such as level spacings and couplings. Also, we do not use a redundant encoding of the computational qubits. A single, two-level system contains the information of a single computational qubit, greatly reducing the complexity of the architecture.

Another novelty introduced by the setup in Fig.~\ref{fig:architecture} is the ability to perform one-shot Toffoli gates~\cite{Zahedinejad2015, Rasmussen202_toffoli, baker2022}. This could, in principle, improve the execution time and the fidelity of many quantum algorithms. The proposed architecture turns out to be very versatile. Indeed, universality could be achieved by substituting the Toffoli gate with two entangling gates as discussed in Appendix~\ref{sec:alternative-designs}. This versatility, together with the potential modularity of this proposal might be useful. From a practical standpoint, the fabrication of a single large loop containing many qubits and the design of global control lines that simultaneously connect all the qubits may prove to be an inefficient and error-prone process. Integrating a hybrid control with different global sources for different subsets of qubits, fabricated on different chiplets, could prove advantageous. 

It is crucial to point out that a major experimental challenge of our architecture arises from spatial inhomogeneities in the fabrication of superconducting qubits and in the always-on ZZ couplings along the conveyor belt, which can affect the fidelity of the required logical operations. However, it has been recently demonstrated~\cite{Aiudi2025} that by employing globally optimized control pulses, it is possible to overcome such disorder and achieve high fidelities.
We stress that all considerations regarding the physical requirements (e.g.~values of the Rabi frequencies, level spacings, ZZ coupling strengths, etc.) for the implementation of our setup remain consistent with those presented in Refs.~\cite{menta2024globally, Aiudi2025}, since the physical building blocks involved are identical.

To achieve fault-tolerant quantum computation, error-correction (EC) protocols are essential. As shown in Fig.~\ref{fig:swap}(b), the number of computational qubits directly coupled to a given one is the same as in conventional 2D arrays, i.e.~four connections. Therefore, in contrast with the proposal of Ref.~\cite{menta2024globally}, our system is equivalent to a 2D processor. This suggests that any standard fault-tolerant procedure can be implemented, such as the famous surface codes~\cite{Fowler2012, google2025}. 
However, recent progress in quantum low-density parity-check (qLDPC) codes~\cite{LDPC1, LDPC2, Panteleev2022, Breuckmann2021} suggest they may be even more suitable for our platform. qLDPC codes maintain sparse parity checks while achieving finite rate and growing distance, enabling higher thresholds and lower overhead compared to surface codes. Importantly, their implementation can benefit from the same local four-qubit connectivity (see, e.g., bivariate bicycle or hypergraph-product codes~\cite{Tillich2014, Kovalev2013, Hastings2021}), which naturally fits our 2D-equivalent architecture.
Finally, although EC schemes for globally-controlled quantum systems have been explored previously~\cite{Bririd_2004, Kay_2005, Kay_2007, Fitzsimons_2007, Fitzsimons_2008}, the design of globally-driven qLDPC protocols specifically tailored to our machine represents a compelling direction for future work.

In conclusion, it should be noted that our setup can be implemented on different physical platforms, including Rydberg atoms~\cite{Saffman_RMP_2010} and semiconductor spin qubits~\cite{Review-Spin-qubits}. 

\section*{Acknowledgments}
We thank M.~Riccardi for useful comments on the draft and discussions. M.P. and V.G. are co-founders and shareholders of Planckian. At the time of their contributions, authors affiliated with Planckian are either employees of Planckian or PhD students collaborating with Planckian.

\appendix

\section{Algebraic properties and pulse constructions for global control}
\label{app:global_control}

In this Appendix we collect technical results underlying the global-control protocol summarized in Sec.~\ref{sec:GCD}. 

\subsection{Control-unitary algebra}

Given $\xi\in\{\chi^{\rm r},\chi^{\times}\}$, the control unitaries
\begin{equation}
\hat{W}_{\xi}(\theta,\bm{n})
=
\prod_{i\in\xi}
\left[
\hat{\openone}_i\otimes\hat{Q}_{\langle i\rangle}
+
\hat{\mathbb{R}}_i(\theta,\bm{n})\otimes\hat{P}_{\langle i\rangle}
\right],
\end{equation}
inherit the group structure of single-qubit $SU(2)$ rotations. In particular, for arbitrary angles $\theta_{1,2}$ and unit vectors $\bm{n}_{1,2}$,
\begin{equation}
\hat{W}_{\xi}(\theta_2,\bm{n}_2)\hat{W}_{\xi}(\theta_1,\bm{n}_1)
=
\hat{W}_{\xi}(\theta_3,\bm{n}_3),
\end{equation}
where $(\theta_3,\bm{n}_3)$ are determined by the corresponding composition rule for
$\hat{\mathbb{R}}_i(\theta,\bm{n})$.
The inverse operation is given by
\begin{equation}
\hat{W}_{\xi}^{-1}(\theta,\bm{n})
=
\hat{W}_{\xi}(-\theta,\bm{n})
=
\hat{W}_{\xi}(\theta,-\bm{n}),
\end{equation}
and the transformations are $4\pi$-periodic in $\theta$.
For $\theta=2\pi$ one obtains a nontrivial phase operation,
\begin{equation}
\hat{W}_{\xi}(2\pi,\bm{n})
=
\prod_{i\in\xi}\left(\hat{Q}_{\langle i\rangle}-\hat{P}_{\langle i\rangle}\right),
\end{equation}
while $\hat{W}_{\xi}(4\pi,\bm{n})=\openone$.

For fixed $\chi\in\{A,B\}$, the regular and crossed contributions commute,
\begin{eqnarray}\nonumber
\hat{W}_{\chi^{\rm r}}(\theta',\bm{n}')
\hat{W}_{\chi^{\times}}(\theta'',\bm{n}'')
=\\
= \hat{W}_{\chi^{\times}}(\theta'',\bm{n}'')
\hat{W}_{\chi^{\rm r}}(\theta',\bm{n}'),
\end{eqnarray}
allowing independent synthesis of operations on the two subsets.

\subsection{Selective control via four-pulse sequences}

A key ingredient for universal control is the ability to selectively address either the regular or the crossed qubits, despite the correlated structure of the elementary gates
$\hat{W}_{\chi}(\theta,\bm{n}_\perp;2\theta,\bm{n}_\perp)$.
This can be achieved using rotations around orthogonal axes in the $xy$ plane.

Let $\bm{n}_\perp$ and $\bm{m}_\perp$ be orthogonal unit vectors with
$\bm{n}_\perp\cdot\bm{m}_\perp=0$.
Using the identity
\begin{equation}
\hat{\mathbb{R}}_i(\pi,-\bm{n}_\perp)
\hat{\mathbb{R}}_i(\theta,\bm{m}_\perp)
\hat{\mathbb{R}}_i(\pi,\bm{n}_\perp)
=
\hat{\mathbb{R}}_i^{-1}(\theta,\bm{m}_\perp),
\end{equation}
one finds that the four-pulse sequence
\begin{equation}
\hat{U}_\chi^{(4)}\hat{U}_\chi^{(3)}\hat{U}_\chi^{(2)}\hat{U}_\chi^{(1)},
\end{equation}
with
\begin{align}
\hat{U}_\chi^{(1)}=\hat{U}_\chi^{(3)}&=\hat{W}_\chi(\theta/4,\bm{n}_\perp;\theta/2,\bm{n}_\perp),\\
\hat{U}_\chi^{(2)}&=\hat{W}_\chi(\pi,\bm{m}_\perp;2\pi,\bm{m}_\perp),\\
\hat{U}_\chi^{(4)}&=\hat{W}_\chi(\pi,-\bm{m}_\perp;2\pi,-\bm{m}_\perp),
\end{align}
implements, in the blockade regime,
\begin{equation}
\hat{U}_\chi^{(4)}\hat{U}_\chi^{(3)}\hat{U}_\chi^{(2)}\hat{U}_\chi^{(1)}
\simeq
\hat{W}_{\chi^{\times}}(\theta,\bm{n}_\perp),
\end{equation}
up to a global phase.
An analogous construction allows one to isolate rotations on the regular subset $\chi^{\rm r}$.
These constructions, combined with Euler-angle decompositions, establish full controllability over both regular and crossed qubits using only global pulses.
Further details and generalizations are provided in Ref.~\cite{menta2024globally}.

\section{Characterizing the exchange operator}\label{sezDYN} 
In this Section we discuss  in  details the action of the exchange operator $\hat{\Pi}_{\rm exc}$ defined in the main text, i.e. 
\begin{equation}\label{swap}
   \hat{\Pi}_{\rm exc} := \hat{\Pi}_B \hat{\Pi}_{A^{\rm r}} \hat{\Pi}_B \hat{\Pi}_{A^{\rm r}} \hat{\Pi}_B \hat{\Pi}_{A^{\rm r}} \hat{\Pi}_B \hat{\Pi}_{A^{\rm r}} = \left(  \hat{\Pi}_B \hat{\Pi}_{A^{\rm r}}\right)^4 \;.
\end{equation}
For convenience, we recall that 
\begin{eqnarray}
\hat{\Pi}_{A^{\rm r}} := \hat{W}_{A^{\rm r}}(\pi, \bm{x})= \prod_{i\in A^{\rm r}} \left[  \hat{\openone}_i \otimes \hat{Q}_{\langle i \rangle}
-i \hat{\sigma}_i^{(x)} \otimes \hat{P}_{\langle i \rangle}  \right]\;
\end{eqnarray} 
and 
\begin{eqnarray} 
\hat{\Pi}_{B}&:=& \hat{W}_{B}(\pi,\bm{x}; \pi ,\bm{x})=
\hat{W}_{B^{\rm r}}(\pi,\bm{x})\hat{W}_{B^{\times}}(\pi,\bm{x})\nonumber \\
&=&  \prod_{i\in B^{\rm r}\cup B^{\rm \times} } \left[  \hat{\openone}_i \otimes \hat{Q}_{\langle i \rangle}
-i \hat{\sigma}_i^{(x)} \otimes \hat{P}_{\langle i \rangle}  \right]\;.
\end{eqnarray}

In particular we are interested in determining how $\hat{\Pi}_{\rm exc}$ acts on the well-formed states
$|\Psi;{\rm FP}\rangle$ and $|\Psi;{\rm PF}\rangle$ that encode the logical information in the model. 
As indicated in Eq.~\eqref{encoding}, being $\Psi_{\vec{k}}$ the probability amplitudes of a $N$-qubit logical state expressed in the computational basis $\{|g\rangle,|e\rangle\}$, 
such states have the form 
\begin{eqnarray} \label{vectors000} 
|\Psi;{\rm FP}\rangle &:=& \sum_{\vec{k}\in \{ g,e\}^N} \Psi_{\vec{k}} \; |\vec{k}; {\rm F P }  ; g \rangle \;, \\
 |\Psi;{\rm PF}\rangle &:=& \sum_{\vec{k}\in \{ g,e\}^N} \Psi_{\vec{k}} \; |\vec{k}; {\rm  PF }  ; g \rangle \;,\end{eqnarray} 
 with  
\begin{eqnarray} \nonumber
|\vec{k}; {\rm F P }  ; g \rangle := |k_1\rangle_{Q_1} |{\rm F} \rangle_{S_1} |k_2\rangle_{Q_2} |{\rm P}\rangle_{S_2}
 |k_3\rangle_{Q_3} |{\rm F}\rangle_{S_3} 
 \cdots \\
 \cdots |k_{N-1}\rangle_{Q_{N-1}}  |{\rm F} \rangle_{S_{N-1}}  |k_N\rangle_{Q_N}  |{\rm P} \rangle_{S_N} \otimes
 |g\rangle_{A^{\times}}\;,
\end{eqnarray} 
where $j\in \{1, \cdots, N\}$, $|{\rm F}\rangle_{S_j}$ and $|{\rm P}\rangle_{S_j}$, define the  ferromagnetic and paramagnetic vectors of the sector $S_j$, respectively. For $|\vec{k}; {\rm P F}  ; g \rangle$ the definition is the same with all $F$ sectors exchanged with $P$ sectors and vice versa.
It is worth remembering that in our setup, all the qubits associated with the IC sites $Q_j$'s are $B$-type
qubits. 
Recall also that each sector $S_j$ is formed by three neighboring sites: 
 the first and last sites host a regular (non-crossed) $A$-type qubit, which we name $A_j^{(1)}$ and $A_j^{(3)}$, and  the second hosts instead 
 a $B$-type qubit which we name  $B_j^{(2)}$, i.e. 
 \begin{eqnarray} 
 S_j := ( A_j^{(1)}, B_j^{(2)},A_j^{(1)})\;. 
 \end{eqnarray} 
 Accordingly we can express the vector $|{\rm F}\rangle_{S_j}$ and $|{\rm P}\rangle_{S_j}$ as 
\begin{eqnarray}
|{\rm F}\rangle_{S_j} &:=& |g\rangle_{A_j^{(1)}}  |g\rangle_{B_j^{(2)}} |g\rangle_{A_j^{(3)}} \;, \\ |{\rm P}\rangle_{S_j} &:=&  |g\rangle_{A_j^{(1)}}  |e\rangle_{B_j^{(2)}} |g\rangle_{A_j^{(3)}} \;.
\end{eqnarray}

\subsection{Evolution of the well-formed state under $\hat{\Pi}_{\rm exc}$} 
To study the action of 
$\hat{\Pi}_{\rm exc}$ on 
the well-formed  states we  shall proceed step-by-step analysing  the role of the individual  pulses that compose it. To simplify the analysis we shall focus on the individual components $|\vec{k}; {\rm F P}  ; g \rangle$ and $|\vec{k}; {\rm P F}  ; g \rangle$, then invoke linearity
to reconstruct  the evolutions of the vectors $|\Psi;{\rm FP}\rangle$ and $|\Psi;{\rm PF}\rangle$.
For this purpose it is useful to
observe the following facts:
\begin{enumerate} 
\item[i)] The Information Carrying (IC) site $Q_j$ admits as neighboring sites
the third element $A_{j-1}^{(3)}$ of the sector $S_{j-1}$, and  the first element
${A_{j+1}^{(1)}}$ of the sector $S_{j+1}$, i.e.
\begin{eqnarray}
A_{j-1}^{(3)} - Q_j - {A_{j+1}^{(1)}}\;.
\end{eqnarray} 
Accordingly the operator $\hat{\Pi}_{B}$ will act on ${Q_j}$  as $-i \hat{\sigma}_{Q_j}^{(x)}$
if and only if both $A_{j-1}^{(3)}$ and $A_{j+1}^{(1)}$ are in the ground configuration. No direct
transformation on $Q_j$ is induced by  $\hat{\Pi}_{A^{\rm r}}$.
Notice that the first three  IC sites $Q_1$, $Q_2$, $Q_3$, also have the
central crossed $A$-type element as  neighboring site.
However,
since this element is always initialized in the ground state $|g\rangle$ and  $\hat{\Pi}_{\rm exc}$ includes no transformations that acting on such term, the central crossed $A$-type element has no role in controlling the evolution of $Q_1$, $Q_2$, $Q_3$ under
the transformation~(\ref{swap}).
\item[ii)]
The central site $B_{j}^{(2)}$ of the sector $S_j$ admits as neighboring sites
the first  $A_{j}^{(1)}$ and the last  $A_{j}^{(3)}$ of the same sector 
\begin{eqnarray}
A_{j}^{(1)} - B_{j}^{(2)} - {A_{j}^{(3)}} \;.
\end{eqnarray} 
Accordingly the operator $\hat{\Pi}_{B}$ will act  on $B_{j}^{(2)}$ as $-i \hat{\sigma}_{B_{j}^{(2)}}^{(x)}$
if and only if both $A_{j}^{(1)}$ and $A_{j}^{(3)}$ are in the ground state. No direct
transformation on $B_{j}^{(2)}$ is induced by  $\hat{\Pi}_{A^{\rm r}}$.
\item[iii)]  The first site $A_{j}^{(1)}$ of the sector $S_j$ admits as neighboring sites
the second element $B_{j}^{(2)}$ of the same sector and the IC site 
$Q_{j-1}$, 
\begin{eqnarray}Q_{j-1}-A_{j}^{(1)} - B_{j}^{(2)}
\;.
\end{eqnarray} 
Accordingly the operator $\hat{\Pi}_{A^{\rm r}}$ will act  on $A_{j}^{(1)}$ as $-i \hat{\sigma}_{A_{j}^{(1)}}^{(x)}$
if and only if both $Q_{j-1}$ and $B_{j}^{(2)}$ are in the ground state. No direct
transformation on $A_{j}^{(1)}$ is induced by  $\hat{\Pi}_{B}$.
\item[iv)]  The third site $A_{j}^{(3)}$ of the sector $S_j$ admits as neighboring sites
the second element $B_{j}^{(2)}$ of the same sector and the IC site 
$Q_{j+1}$, 
\begin{eqnarray}B_{j}^{(2)} -A_{j}^{(3)} - Q_{j+1}
\;.
\end{eqnarray} 
Accordingly the operator $\hat{\Pi}_{A^{\rm r}}$ will act  on $A_{j}^{(3)}$ as $-i \hat{\sigma}_{A_{j}^{(3)}}^{(x)}$
if and only if both $Q_{j+1}$ and $B_{j}^{(2)}$ are in the ground state. No direct
transformation on $A_{j}^{(3)}$ is induced by  $\hat{\Pi}_{B}$.
\end{enumerate} 
Equipped with the above observations we can now proceed with the step-by-step analysis of the evolution induced by the unitary operator~(\ref{swap}). For clarity, here we show only the action of the first two pulses. The action of the entire sequence can be derived easily using the above stated facts.

\begin{enumerate} 
\item {\bf First pulse ($\hat{\Pi}_{A^{\rm r}}$):} This transformation acts directly only on the regular
$A$-type sites. In our model they are only present inside the sectors $S_j$ which in the input
state are either in a paramagnetic ($|{\rm P}\rangle_{S_j}$) or a ferromagnetic ($|{\rm F}\rangle_{S_j}$)
configuration. 
Notice that the presence of the $|e\rangle_{B_j^{(2)}}$ element in the formula of $|{\rm P}\rangle_{S_j}$ prevents 
$\hat{\Pi}_{A^{\rm r}}$ from modifying  such part of the vectors $|\vec{k}; {\rm F P}  ; g \rangle$ and $|\vec{k}; {\rm P F}  ; g \rangle$, i.e.
\begin{eqnarray} 
\hat{\Pi}_{A^{\rm r}}  |{\rm P}\rangle_{S_j} =  |{\rm P}\rangle_{S_j} \;. \label{prima}
\end{eqnarray} 
To evaluate the effect of  $\hat{\Pi}_{A^{\rm r}}$ on the components $|{\rm F}\rangle_{S_j}$, we need to 
take into account the state of the neighboring IC sites that are directly connected to them, i.e. the vectors 
$|k_j\rangle_{Q_j}$ and $|k_{j+1}\rangle_{Q_{j+1}}$.
 In this case we have 
\begin{widetext}
\begin{eqnarray} 
\hat{\Pi}_{A^{\rm r}} |k_j\rangle_{Q_j} |{\rm F}\rangle_{S_j} |k_{j+1}\rangle_{Q_{j+1}}&=&
\hat{\Pi}_{A^{\rm r}} |k_j\rangle_{Q_j} |g\rangle_{A_j^{(1)}}  |g\rangle_{B_j^{(2)}} |g\rangle_{A_j^{(3)}}   |k_{j+1}\rangle_{Q_{j+1}}\nonumber \\
&=&\left\{ \begin{array}{rrr}
 (-i)^2 |g\rangle_{Q_j} |e\rangle_{A_j^{(1)}}  |g\rangle_{B_j^{(2)}} |e\rangle_{A_j^{(3)}}   |g\rangle_{Q_{j+1}}
 && \mbox{for $(k_j,k_{j+1}) = (g,g)$,} \\
(  -i )\; |g\rangle_{Q_j} |e\rangle_{A_j^{(1)}}  |g\rangle_{B_j^{(2)}} |g\rangle_{A_j^{(3)}}   |e\rangle_{Q_{j+1}}
   && \mbox{for $(k_j,k_{j+1}) = (g,e)$,}\\
(   -i)\;  |e\rangle_{Q_j} |g\rangle_{A_j^{(1)}}  |g\rangle_{B_j^{(2)}} |e\rangle_{A_j^{(3)}}   |g\rangle_{Q_{j+1}}
    && \mbox{for $(k_j,k_{j+1}) = (e,g)$,}\\
   |e\rangle_{Q_j} |g\rangle_{A_j^{(1)}}  |g\rangle_{B_j^{(2)}} |g\rangle_{A_j^{(3)}}   |e\rangle_{Q_{j+1}}
    && \mbox{for $(k_j,k_{j+1}) = (e,e)$.} 
\end{array} 
\right. \label{seconda} 
\end{eqnarray} 
\end{widetext}
Eqs.~(\ref{prima}) and (\ref{seconda}) together determine the evolution of 
$|\vec{k}; {\rm F P }  ; g \rangle$ and $|\vec{k}; {\rm PF }  ; g \rangle$, and hence of  
$|\Psi;{\rm FP}\rangle$ and $|\Psi;{\rm FP}\rangle$, under the action of the first component $\hat{\Pi}_{A^{\rm r}}$ of $\hat{\Pi}_{\rm exc}$. 
\item {\bf Second pulse ($\hat{\Pi}_{B}$):} Let us next apply $\hat{\Pi}_{B}$ on the transformed vectors
which emerge from the application of the first $\hat{\Pi}_{A^{\rm r}}$ operator. In this case, only the IC site $Q_j$ and the central elements of sectors $S_j$ are directly affected by the evolution. 

To begin with notice from Eq.~\eqref{prima} it follows
\begin{eqnarray} 
\hat{\Pi}_{B}  \hat{\Pi}_{A^{\rm r}}  |{\rm P}\rangle_{S_j} = \hat{\Pi}_{B}  |{\rm P}\rangle_{S_j}=
(-i)|{\rm F}\rangle_{S_j}\;. 
\label{terza'}
\end{eqnarray} 
where we use the fact that the $B$-type qubit inside paramagnetic sector
gets flipped because it is surrounded by  $|g\rangle$ qubits. 
To evaluate the action of $\hat{\Pi}_{B}$ on the vectors (\ref{seconda}) we need to include their first neighboring sites: the third $A$-type qubit of the sector $S_{j-1}$ (i.e. $A_{j-1}^{(3)}$)  and the first $A$-type qubit of the sector $S_{j+1}$ (i.e. $A_{j+1}^{(1)}$).
By construction, these qubits were originally assigned to paramagnetic phases and, therefore (due to Eq.~(\ref{prima})), have remained in the state $|g\rangle$ from which they started. Consequently, they do not interfere with the action of $\hat{\Pi}_{B}$ on the $B$-type qubits of the vectors (\ref{seconda}); the only control being exerted by $A_{j}^{(1)}$ and $A_{j}^{(3)}$.
Therefore we obtain 
\begin{widetext}
\begin{eqnarray}  
\hat{\Pi}_{B} \hat{\Pi}_{A^{\rm r}} |k_j\rangle_{Q_j} |{\rm F}\rangle_{S_j} |k_{j+1}\rangle_{Q_{j+1}} &=&\left\{ \begin{array}{rrr}
 (-i)^2 |g\rangle_{Q_j} |e\rangle_{A_j^{(1)}}  |g\rangle_{B_j^{(2)}} |e\rangle_{A_j^{(3)}}   |g\rangle_{Q_{j+1}}
 && \mbox{for $(k_j,k_{j+1}) = (g,g)$,} \\
  (-i)^2 |g\rangle_{Q_j} |e\rangle_{A_j^{(1)}}  |g\rangle_{B_j^{(2)}} |g\rangle_{A_j^{(3)}}   |g\rangle_{Q_{j+1}}
   && \mbox{for $(k_j,k_{j+1}) = (g,e)$,}\\
  (-i)^2 |g\rangle_{Q_j} |g\rangle_{A_j^{(1)}}  |g\rangle_{B_j^{(2)}} |e\rangle_{A_j^{(3)}}   |g\rangle_{Q_{j+1}}
    && \mbox{for $(k_j,k_{j+1}) = (e,g)$,}\\
(-i)^3 |g\rangle_{Q_j} |g\rangle_{A_j^{(1)}}  |e\rangle_{B_j^{(2)}} |g\rangle_{A_j^{(3)}}   |g\rangle_{Q_{j+1}}
    && \mbox{for $(k_j,k_{j+1}) = (e,e)$.} 
\end{array} \label{seconda'} 
\right.
\end{eqnarray} 
\end{widetext}
Notice that, according to the above expression, after the action of 
$\hat{\Pi}_{B}  \hat{\Pi}_{A^{\rm r}}$, all the IC qubits $Q_j$ are in the ground state, so they never block the action of the next $\hat{\Pi}_{A^{\rm r}}$ pulse. 
As usual, combining (\ref{terza'}) and (\ref{seconda'}) we obtain the evolution of 
$|\vec{k}; {\rm F P }  ; g \rangle$ and $|\vec{k}; {\rm PF }  ; g \rangle$, and hence of  
$|\Psi;{\rm FP}\rangle$ and $|\Psi;{\rm FP}\rangle$, under the action of the first two  components $\hat{\Pi}_{B}  \hat{\Pi}_{A^{\rm r}}$ of $\hat{\Pi}_{\rm exc}$. 
\end{enumerate}

The action of the following six pulses can be obtained in a similar fashion. In particular, one finds that whenever two consecutive IC sites of the system 
 are separated by a ferromagnetic region $|{\rm F}\rangle_{S_j}$, their internal state are
  swapped by $\hat{\Pi}_{\rm exc}$, while
 $|{\rm F}\rangle_{S_j}$  gets replaced by the paramagnetic region  $|{\rm P}\rangle_{S_j}$.
Similarly, under $\hat{\Pi}_{\rm exc}$, the components $|{\rm P}\rangle_{S_j}$ gets transformed into $|{\rm F}\rangle_{S_j}$.
For example, for $|\vec{k}; {\rm F P }  ; g \rangle$, we can write 
\begin{eqnarray} 
\hat{\Pi}_{\rm exc} |\vec{k}; {\rm F P }  ; g \rangle = \hat{U}^{\rm swap}_{Q_1 Q_2}\cdots
  \hat{U}^{\rm swap}_{Q_{N-1} Q_N}  |\vec{k}; {\rm PF }  ; g \rangle \;,\label{vectors11}
\end{eqnarray} 
where the swap operations acts only on the IC qubits. 
 A final  simplification arises by observing that the sequence of the swapping gates
 $\hat{U}^{\rm swap}_{Q_1 Q_2} \hat{U}^{\rm swap}_{Q_3 Q_4}\cdots
  \hat{U}^{\rm swap}_{Q_{N-1} Q_N}$ corresponds to  a single-step, clock-wise
   rotation on the internal state of the IC qubits $Q_j$ with $j$ odd, and a 
   single-step, anti-clock-wise
   rotation on the internal state of the IC qubits $Q_j$ with $j$ even.  
At the level of the vectors (\ref{vectors000}) this leads to the identity 
  \begin{eqnarray} 
  \hat{\Pi}_{\rm exc} |\Psi; {\rm FP}\rangle = |\Psi^{(1)}_{{\circlearrowright}_o,{\circlearrowleft}_e}; {\rm PF}\rangle \;, \label{evol} 
\end{eqnarray} 
 reported in the main text. A similar simplification  for $|\vec{k}; {\rm P F}  ; g \rangle$ leads to 

 \begin{eqnarray}
     \hat{\Pi}_{\rm exc} |\Psi; {\rm PF}\rangle = |\Psi^{(1)}_{{\circlearrowleft}_o,{\circlearrowright}_e}; {\rm FP}\rangle\;,
 \end{eqnarray}

\subsection{Concatenation rule} 
An important aspect of the evolution~(\ref{evol}), is that iterative applications of 
 $\hat{\Pi}_{\rm exc}$ on a well-formed state do not cancel out.
For example, for $|\Psi; {\rm FP}\rangle$ we have 
   \begin{eqnarray} 
\left( \hat{\Pi}_{\rm exc}\right)^{2}  |\Psi; {\rm FP}\rangle = |\Psi^{(2)}_{{\circlearrowright}_o,{\circlearrowleft}_e}; {\rm FP}\rangle \;, \label{evol2} 
\end{eqnarray} 
where now the logical state of $|\Psi^{(2)}_{{\circlearrowright}_o,{\circlearrowleft}_e}; {\rm FP}\rangle$ (resp.
$|\Psi^{(2)}_{{\circlearrowleft}_o,{\circlearrowright}_e}; {\rm PF}\rangle$), is obtained by
applying a two-step, clock-wise
   rotation on the internal state of the active IC sites $Q_j$ with $j$ odd (even), and a 
   two-step, anti-clock-wise
   rotation on the internal state of the IC qubits $Q_j$ with $j$ even (odd). 
   The reason for this is that, due to  Eq.~(\ref{evol}), the odd and even IC sites of 
   $\hat{\Pi}_{\rm exc} |\Psi; {\rm FP}\rangle$ have been exchanged. Therefore when we act 
   with 
   $\hat{\Pi}_{\rm exc}$ on such configuration, despite the fact that the system is now in a ${\rm PF}$ well-formed vector, the net effect is still to induce an extra  single-step clock-wise rotation on the IC qubits
   with $j$ odd, and an extra  single-step anti-clock-wise rotation on the IC qubits
   with $j$ even. 
Building upon this for  $\ell$ integer, it then follows that we can write 
   \begin{eqnarray} 
 \left( \hat{\Pi}_{\rm exc}\right)^{\ell}  |\Psi; {\rm FP}\rangle = \left\{
 \begin{array}{ll} |\Psi^{(\ell)}_{{\circlearrowright}_o,{\circlearrowleft}_e}; {\rm PF}\rangle & \mbox{for $\ell$ odd,}\\\\ |\Psi^{(\ell)}_{{\circlearrowright}_o,{\circlearrowleft}_e}; {\rm FP}\rangle & \mbox{for $\ell$ even,}
 \end{array} 
\right. \label{ellrule}
\end{eqnarray} 
Observe that of course for $\ell=N$, the unitary $\left( \hat{\Pi}_{\rm exc}\right)^{\ell}$ acts as the identity transformation, i.e.
   \begin{eqnarray} 
 \left( \hat{\Pi}_{\rm exc}\right)^{N}  |\Psi; {\rm FP}\rangle =  |\Psi; {\rm FP}\rangle\;,
 \end{eqnarray} 
due to the fact that $\Psi^{(N)}_{{\circlearrowright}_o,{\circlearrowleft}_e}=\Psi^{(N)}_{{\circlearrowleft}_o,{\circlearrowright}_e}=\Psi$. 
Accordingly the action of $\left( \hat{\Pi}_{\rm exc}\right)^{\ell}$  is “inverted" by 
$\left( \hat{\Pi}_{\rm exc}\right)^{N-\ell}$. An alternative way to realize such effect is to use
the transformation  
\begin{eqnarray} \hat{\Pi}'_{\rm exc}:=\hat{\Pi}_B\hat{\Pi}_{\rm exc} \hat{\Pi}_B\;, \end{eqnarray} 
that effectively acts as the inverse of $\hat{\Pi}_{\rm exc}$ when operating on well-formed states (the proof of this assertion follows from the same derivation presented here). 
A direct consequence of (\ref{ellrule}) 
is that, given any target values $j,j'\in \{ 1,\cdots, N\}$, we can use our control pulses to induce a transformation that
brings the input state of the $j$-th IC site of any well-formed state into the $j'$-th IC site.

\section{Universal Quantum Computing}\label{secENC}

Here we prove that, using the encoding provided by well-formed states
$|\Psi; {\rm FP}\rangle$ and $|\Psi; {\rm PF}\rangle$ the setup allows for universal QC. 
The starting point of the analysis are the following facts:
\begin{itemize} 
\item[1)] Using  iterative application of the transformation~$\hat{\Pi}_{\rm exc}$ we can induce
cyclic rotations among the IC sites. 
\item[2)]  We can realize all possible single qubit transformation on any IC sites of the model.
\item[3)]  We can realize the Toffoli gate $\hat{T}_{1,3\rightarrow 2}$ which has $Q_1$ and $Q_3$ as the controller qubits and $Q_2$ as the controlled one. 
\end{itemize}  
Thanks to these properties, universality can be proved by simply showing that one can induce individual,  two-qubit swaps $\hat{U}^{\rm swap}_{Q_j Q_{j'}}$ among all possible couples $\{ Q_j,  Q_{j'}\}$ of IC qubits. Indeed  
if we attain such a task, then we can convert $\hat{T}_{1,3\rightarrow 2}$  in an arbitrary Toffoli transformation
$\hat{T}_{j_1,j_2\rightarrow j_3}$ that couples each possible triple $\{ Q_{j_1},  Q_{j_2}, Q_{j_3}\}$ of the system. 
Then we can invoke the fact that universal QC is granted as soon as you can induce arbitrary single-gate transformations (point 2) of the above list, and arbitrary Toffoli gates. 

\subsection{Inducing all possible swap transformations} 
Here, we show that using the properties 1), 2), and 3), we can generate all the individual two-body swap
gates   among the IC sites of the model. 
This problem can be mapped into a graph problem. The idea is to represent each IC site
of the setup as individual vertex of a graph and to draw  an edge between two of them if and only if there is
a sequence of operations that, using the properties 1), 2), and 3), allows us to implement the swap gate between the corresponding IC elements. 
In this context, proving the thesis means being able to show that in the end, the  graph is fully connected.  

Let us start from some preliminary observations. 
Given $a,b,c$ qubits,  the CNOT gate $\hat{U}^{\rm cnot}_{a\rightarrow c}$   with $a$  being the controller and $c$ the controlled element, can be realized by concatenating two  Toffoli transformations $\hat{T}_{a,b\rightarrow c}$ plus two local operations
on the $b$  qubit, i.e.
\begin{eqnarray} 
\hat{U}^{\rm cnot}_{a\rightarrow c} = \hat{T}_{a,b\rightarrow c}\;  \hat{\sigma}_b^{(x)} \; \hat{T}_{a,b\rightarrow c}
\hat{\sigma}_b^{(x)}\;.
\end{eqnarray} 
This can hence be transformed into a CNOT gate $\hat{U}^{\rm cnot}_{c\rightarrow a}$   where $c$  is the controller and $a$ the controlled qubit, by using extra local operations on $a$ and $c$, i.e. 
\begin{eqnarray} \nonumber
&&\hat{U}^{\rm cnot}_{c\rightarrow a} =  \hat{H}_a  \hat{H}_c \hat{U}^{\rm cnot}_{a\rightarrow c} \hat{H}_a  \hat{H}_c =\\ 
&&=
\hat{H}_a  \hat{H}_c\left( \hat{T}_{a,b\rightarrow c}\;  \hat{\sigma}_b^{(x)} \; \hat{T}_{a,b\rightarrow c} 
\hat{\sigma}_b^{(x)}\right)  \hat{H}_a  \hat{H}_c\;, 
\end{eqnarray} 

with $\hat{H}_{a}$  and $\hat{H}_{c}$ being Hadamard gates on $a$ and $c$ qubits, respectively~\cite{Nielsen2010}. 
Concatenating  $\hat{U}^{\rm cnot}_{a\rightarrow c}$ and $\hat{U}^{\rm cnot}_{c\rightarrow a}$ we can then realize a 
swap gate among $a$ and $c$, i.e. 
\begin{widetext}
\begin{eqnarray} 
\hat{U}^{\rm swap}_{ac} &=&  \hat{U}^{\rm cnot}_{a\rightarrow c}\hat{U}^{\rm cnot}_{c\rightarrow a}  \hat{U}^{\rm cnot}_{a\rightarrow c}\nonumber \\
&=&\left(\hat{T}_{a,b\rightarrow c}\;  \hat{\sigma}_b^{(x)} \; \hat{T}_{a,b\rightarrow c} 
\hat{\sigma}_b^{(x)}\right)
 \hat{H}_a  \hat{H}_c\left( \hat{T}_{a,b\rightarrow c}\;  \hat{\sigma}_b^{(x)} \; \hat{T}_{a,b\rightarrow c} 
\hat{\sigma}_b^{(x)}\right) \hat{H}_a  \hat{H}_c \left(\hat{T}_{a,b\rightarrow c}\;  \hat{\sigma}_b^{(x)} \; \hat{T}_{a,b\rightarrow c} 
\hat{\sigma}_b^{(x)}\right)\;. 
\end{eqnarray} 
\end{widetext}
 Since $\hat{T}_{a,b\rightarrow c}$ is symmetric with respect to the exchange between $a$ and $b$, 
 the previous analysis can also be used to show that using $\hat{T}_{a,b\rightarrow c}$ and 
 local operations, also the swap gate $\hat{U}^{\rm swap}_{bc}$ is attainable. From that we can finally
 construct $\hat{U}^{\rm swap}_{ab}$ by simple concatenation of the previous two, i.e.
 \begin{eqnarray} \label{rule2} 
\hat{U}^{\rm swap}_{ab} &=& \hat{U}^{\rm swap}_{ac}\hat{U}^{\rm swap}_{bc}
 \hat{U}^{\rm swap}_{ac}\;. 
\end{eqnarray} 
Accordingly we can say that 
 \begin{eqnarray} \label{rule1} 
\hat{T}_{a,b\rightarrow c}  \  + \  \mbox{local ops}\  \mapsto \  \{ \hat{U}^{\rm swap}_{ac},
\hat{U}^{\rm swap}_{bc},
 \hat{U}^{\rm swap}_{ab}\} \;. 
\end{eqnarray} 
Thanks to this result, from the properties 2) and 3) we can conclude that 
 in our graph problem we can draw at least   three edges among the sites $Q_1$, $Q_2$ and $Q_3$. 
 Let us now use $\hat{\Pi}_{\rm exc}^{(\ell)}$ to induce a rotation of these sites. Recalling that
 the even and odd sites of the model counter-propagate, we can ensure that after this transformation,
the sites $(Q_1, Q_2, Q_3)$ are mapped (for instance) to $(Q_2, Q_1, Q_4)$. In conjunction with $\hat{T}_{1,3\rightarrow 2}$ this enables us to realize the Toffoli gate $\hat{T}_{2,1\rightarrow 4}$. Invoking (\ref{rule1}), this implies that
 we can also acquire all the swap gates between $Q_1$, $Q_2$ and $Q_4$. Additionally, we can have
 an extra swap between $Q_3$ and $Q_4$ as a consequence of the composition rule (\ref{rule2}).
This implies that in our problem, the first four sites are fully connected. 
Applying further rotations will increase the number of edges. To show that in the end, we can fully connect the entire
graph observe that, if we started from a ${\rm FP}$ well-formed state,  iterative applications of $\hat{\Pi}_{\rm exc}^{(\ell)}$ will force the
odd sites to rotate clock-wise. In particular, after an even number of applications of 
$\hat{\Pi}_{\rm exc}$, in the position originally occupied by $Q_1$ and $Q_3$, we will have a generic couple of consecutive odd elements 
$Q_{2j-1}$ and $Q_{2j+1}$. Accordingly, using  $\hat{T}_{1,3\rightarrow 2}$, we can now
 generate  $\hat{T}_{2j-1,2j+1\rightarrow 2j'}$, with $2j'$ being some even index that is not important to determine at this level. Hence, using  (\ref{rule1}) and  (\ref{rule2}) we can conclude that we will be able to
connect  the two odd sites $Q_{2j-1}$ and $Q_{2j+1}$ with an edge.
Since $j$ is arbitrary, this implies that in our model, the subgraph associated with the
 odd sites is fully connected. A similar argument can be used to conclude that also the subgraph 
 associated with the
 even sites is also fully connected. Notice also that the odd and even subgraphs are connected by at least one edge
 (e.g. the one associated with the swap gate between $Q_1$ and $Q_2$). Invoking the percolation 
 property (\ref{rule2}), this single connection can then be used to easily verify that {\it any} other edges connecting the two subgraphs is also achievable, concluding the thesis.

\section{Initialization and Read-out}\label{sec:init-readout}
\subsection{Initialization}
As mentioned in the main text we can initialize the system into a well-formed state starting from a configuration where all the sites (including the IC ones) are in the ground state. 
For this purpose it is indeed sufficient to use the control $V_{\rm init}(t)$ to induce a $\pi$-pulse that brings
the associated qubits from $|g\rangle$ to $|e\rangle$. Since such elements are internal $B$-type qubits of 
alternating sectors of the device, this will force those sectors to assume a paramagnetic phase $|{\rm P}
\rangle$. In the case of  the scheme of Fig.~1 of the main text this, will produce a ${\rm FP}$ well-formed state 
\begin{eqnarray} \nonumber
|\Psi_0; {\rm F P }  ; g \rangle := |g\rangle_{Q_1} |{\rm F} \rangle_{S_1} |g\rangle_{Q_2} |{\rm P}\rangle_{S_2}
 |g\rangle_{Q_3} |{\rm F}\rangle_{S_3} 
 \cdots \\
 \cdots |g\rangle_{Q_{N-1}}  |{\rm F} \rangle_{S_{N-1}}  |g\rangle_{Q_N}  |{\rm P} \rangle_{S_N} \otimes
 |g\rangle_{A^{\times}}\;,
\end{eqnarray} 
with $|\Psi_0\rangle$ being the logical state where all the IC qubits are in the ground state. 

\subsection{Read-out}
The read-out procedure is more tricky. Indeed, since the read-out of multiple qubits can be done only locally on each qubit, we cannot perform the read-out directly on the quantum processing unit (i.e. the architecture design of Fig.~1 of the main text). The idea is to use a register, which is simply an additional conveyor belt-like wire with no $A$-type crossed qubit inside, but still with a single $B$-type crossed qubit to perform single-qubit gate on it. Such additional area may be used to host all the quantum information moving from the processing unit, at the moment of the read-out. Since this additional read-out area is not involved in the computation, we are “allowed" to use local control lines on the $Q^{\rm read-out}_1, \ldots , Q^{\rm read-out}_N$ elements of the additional setup. The two conveyor belt systems (the processing unit and the read-out register) are coupled via an additional $C$-type qubit (controlled by an additional source $V_C(t)$), which is in turn ZZ coupled to the two $B$-type crossed qubits of the processing unit and read-out area, respectively. Such $C$-type qubit allows to implement a two-qubit operation between the qubits of the two islands, the logical state of processing unit and the “empty" qubit of the read-out register.
\\ \\
Once the computation is over, let us call the final well-formed state $|\Psi'; {\rm F P }  ; g \rangle$, the protocol for the read-out procedure can be schematised as follows
\begin{itemize}
    \item[\textbf{a)}] {\it Initialization of the read-out register}. \\ \\
    The whole state of the read-out area is initialized as explained above for the processing unit area.
    \item[\textbf{b)}] {\it Transfer of the quantum information into the read-out area}. \\ \\
    Combining single-qubit local operations on the two $B$-type crossed qubits of the two islands, with two-qubit gates performed on the $C$-type inter qubit which couples the two islands, a SWAP gate can be performed~\cite{Nielsen2010}. This allows us to transfer (swap) a computational qubit form the processing unit the read-out area (and vice-versa).
    \item[\textbf{c)}] {\it Total resetting of the processing unit area and transfer completion.} \\ \\
    As explained in the main text, through sequences of global pulses $\hat{\Pi}_{\rm exc}$, we are able to bring each logical state in the position corresponding to the $B$-type crossed element of the processing unit area. Once done, we repeat the step \textbf{b)} and subsequently \textbf{c)} until the final well-formed state of the processing unit area will be completely moved into the read-out area and vice-versa: $|\Psi'; {\rm F P }  ; g \rangle \leftrightarrow |\Psi_0; {\rm F P }  ; g \rangle_{\rm read-out}$.
\end{itemize}
Upon completion of the aforementioned step-by-step procedure, each logical state within the read-out area can be measured independently. The objective of this Section is to emphasize the potential realization of a separate globally driven read-out area, equipped with $N$ local control lines (probes) used to make measurements on each $Q_j$ computational qubit.

\section{Alternative designs}\label{sec:alternative-designs}
In this Section we discuss two alternative ways of implementing multi-qubit gates, thus allowing to perform a universal quantum computation in our scheme. Specifically, these alternative implementations are based on two-qubit gates~\cite{menta2024globally} (instead of a single Toffoli gate), which can be performed by simply cutting one of the three connections to the central crossed $A$-type element in Fig.~\ref{fig:architecture}. In this case, (regular, crossed or double-crossed) qubits inside the conveyor-belt QC will mediate a two-qubit CZ gate, instead of a three-qubit CCZ gate. 
\subsection{Proof of universality}
Suppose we are able to perform two-qubit gates between the computational qubits of two pairs of IC site, say $(Q_1,Q_3)$ and $(Q_1,Q_2)$. Two possible ways of independently decide which of the two pairs we are controlling will be presented later on. The combination of the CZ gate with single qubit operations allows to implement any possible two-qubit gate, including the swap gates $\hat{U}^{\rm swap}_{Q_1Q_2}$ and $\hat{U}^{\rm swap}_{Q_1Q_3}$. Thanks to Eq.~\eqref{rule2}, we are also able to implement $\hat{U}^{\rm swap}_{Q_2Q_3}$, making the graph formed by the three IC sites fully connected, which is exactly what we proved for the case of the Toffoli gate. The same proof for the universality presented in Section~\ref{secENC} can thus be applied to the present case.

\subsection{Independent two-qubit gates}
\begin{figure*}[tbp]
\centering
\includegraphics[scale=0.25]{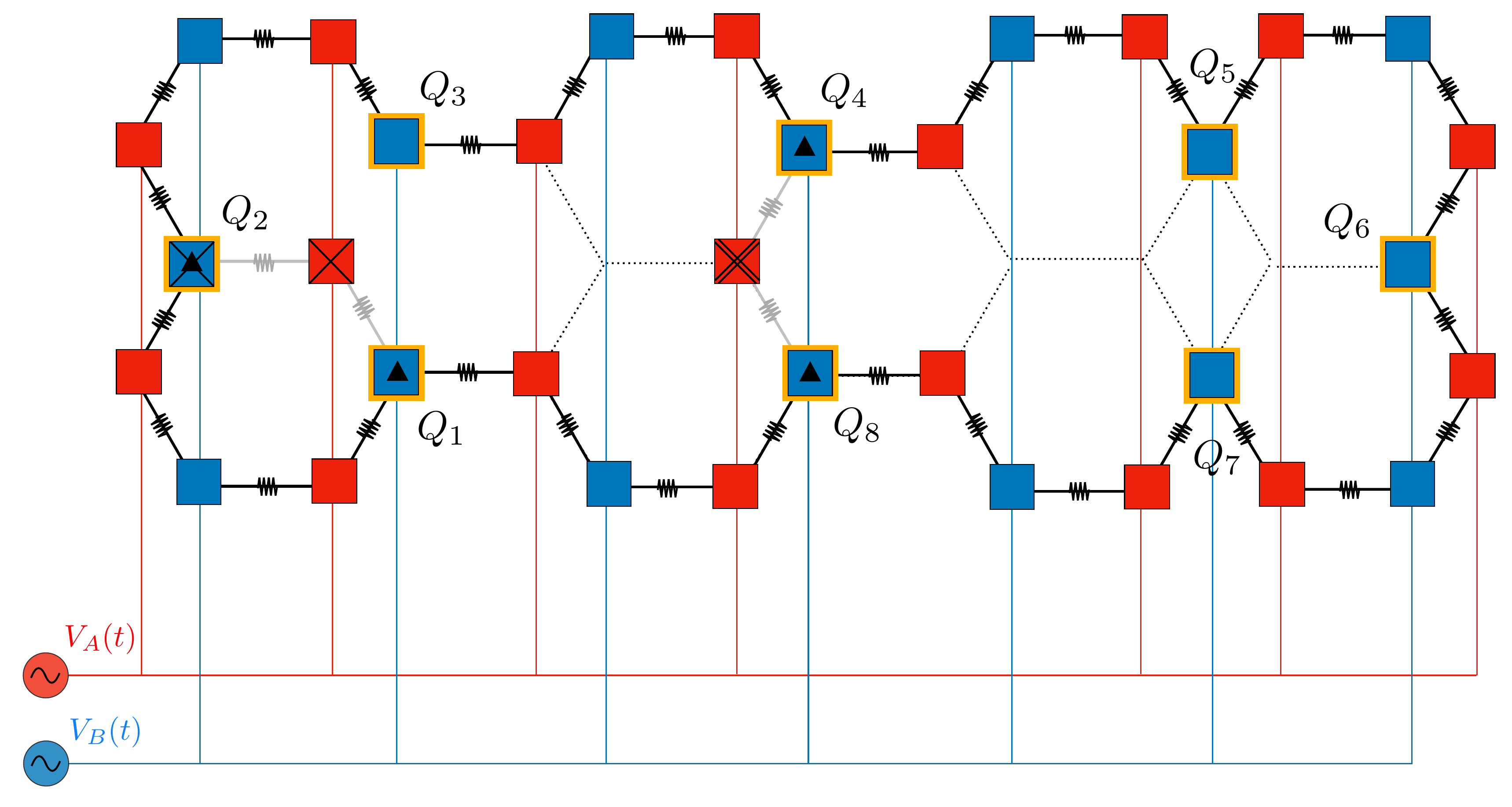}
\caption{\label{fig:alternative1} Alternative design with respect to the conveyor-belt quantum computer of the main text. Two two-qubit gates are present in the setup to ensure the universality of the computation. In order to control them independently we employ a third Rabi frequency (A-type double-crossed qubit), see Eq.~\eqref{double-crossed}.}
\end{figure*}

Referring to the discussion above, how do we select which pair of qubits, $(Q_1,Q_3)$ or $(Q_1,Q_2)$, we are acting on? It does not suffice to connect such qubits via two $A$-type crossed elements. Indeed, since the control is global, the pulse $\hat{\mathcal{Z}}_{A^{\times}}:=\hat{W}_{A^{\times}}(2\pi, \bm{n})$ (see main text) would perform a CZ gate simultaneously on the two pairs, and the control would not be independent. A possible way of breaking this symmetry is to employ an additional control line. In this alternative design, the two qubits placed inside the loop belong to a third species named $C$, and are controlled by a third control line $V_C(t)$. To independently control them, we require one of the two, say the one connecting $Q_1$ and $Q_2$, to be crossed (double Rabi frequency) \cite{menta2024globally}. Then, the pulse $\hat{W}_{C^{\rm r}}(2\pi, \bm{n})$ implements a CZ gate between the qubits at sites $Q_1$ and $Q_3$, while the pulse $\hat{W}_{C^{\times}}(2\pi, \bm{n})$ implements a CZ gate between the qubits at sites $Q_1$ and $Q_3$.\\
However it turns out that it is also possible to maintain the two species scheme. Indeed it can be proved~\cite{menta2025building} that, given three qubits of the same species $\chi$, with Rabi frequencies equal to $\Omega_\chi$, $2\Omega_\chi$ and $4\Omega_\chi$, an independent control of the three can be performed, i.e.~referring to the qubits with frequency $4\Omega_\chi$ as double-crossed ($\mathbb{X}$) elements, we can perform operations of the form
\begin{eqnarray}
    &&\hat{W}_{\chi}(\theta',\bm{n}'; \theta'',\bm{n}''; \theta''',\bm{n}''')
:=  \\\nonumber
&&:= \hat{W}_{\chi^{\rm r}}(\theta', \bm{n}') \hat{W}_{\chi^{\times}}(\theta'', \bm{n}'') \hat{W}_{\chi^{\mathbb{X}}}(\theta''', \bm{n}''')  \;,
\label{double-crossed}
\end{eqnarray}
where $\hat{W}_{\chi^{\rm r}}(\theta', \bm{n}')$, $\hat{W}_{\chi^{\times}}(\theta'', \bm{n}'')$ and $\hat{W}_{\chi^{\mathbb{X}}}(\theta''', \bm{n}''')$ apply only to the regular, crossed and double-crossed qubits, respectively. In the case of Fig.~\ref{fig:alternative1}, there is only one crossed element of type $A$.  We stress that all angles and vectors in the above equation are completely independent. Following the discussion of the previous implementation, it is straightforward to prove that, by making one of the two qubits inside the loop crossed and the other double-crossed, we achieve a universal computation. \\ \\

\begin{figure*}[t]
\centering
\includegraphics[scale=0.25]{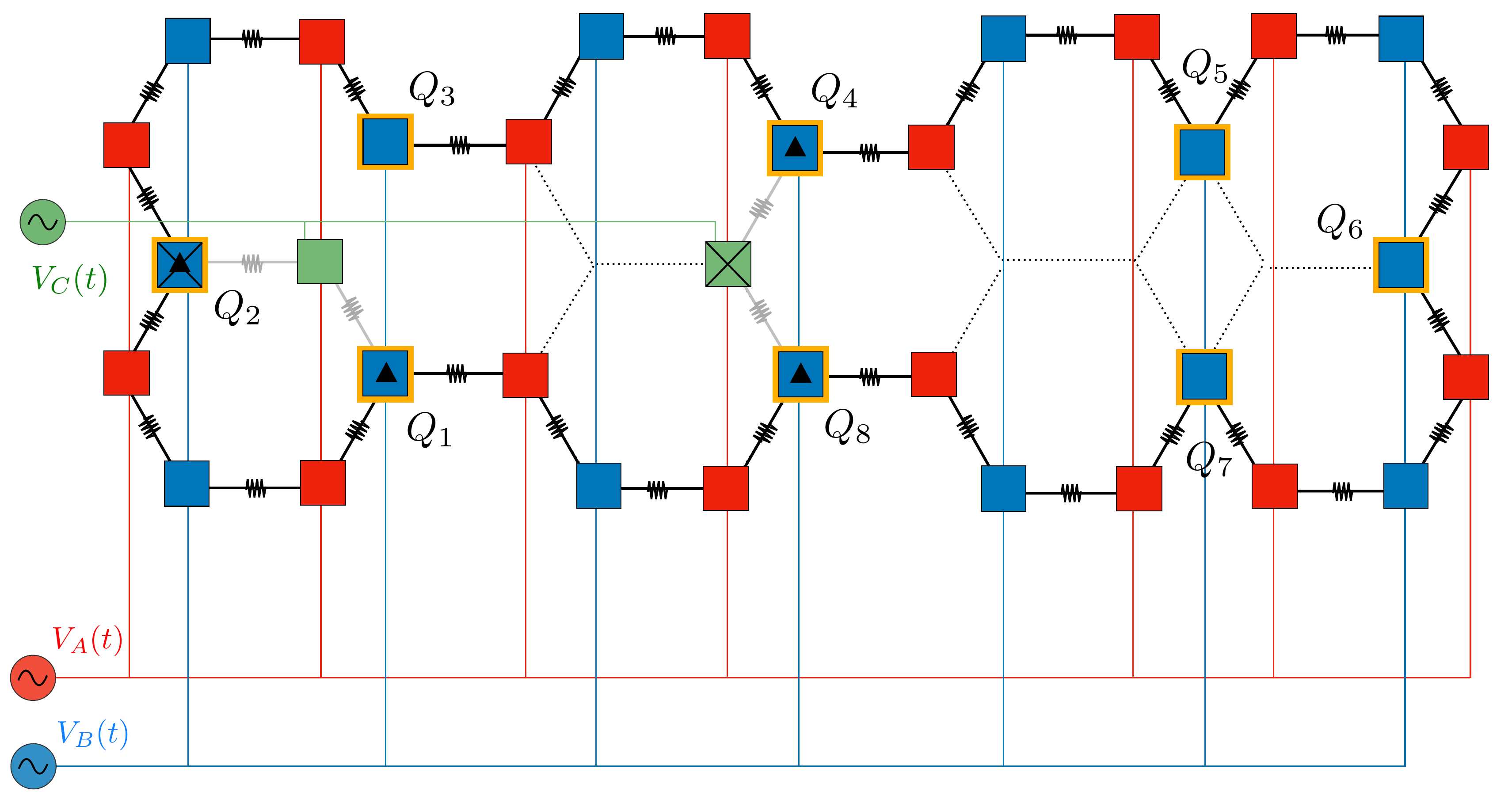}
\caption{\label{fig:alternative2} Alternative design with respect to the conveyor-belt quantum computer of the main text. Two two-qubit gates are present in the setup to ensure the universality of the computation. The two qubits which mediate the two-qubit gates are driven by a third control source $V_C(t)$, i.e. they belong to a third, namely $C$, species (green elements): in order to control them independently the Rabi frequency must be different (one of them is regular, the other one is crossed), see Eq.~\eqref{crossed_C}.}
\end{figure*}

Another equivalent alternative design is depicted in Fig.~\ref{fig:alternative2}, where two $C$-type qubits (one regular and the other one crossed) are introduced instead of the $A$-type qubits of Fig.~\ref{fig:alternative1}.

The $C$-type qubits, consisting of one regular and one crossed qubit, can be controlled independently, ensuring the universality of the computation. In other words, we can perform an evolution
\begin{equation}
\hat{W}_{C}(\theta',\bm{n}'; \theta'',\bm{n}'')
:= \hat{W}_{C^{\rm r}}(\theta', \bm{n}') \hat{W}_{C^{\times}}(\theta'', \bm{n}'') 
\label{crossed_C}
\end{equation}
which allows operations on the two $C$-type elements, thereby mediating two independent two-qubit gates involving computational qubits of different parity.

Finally, we stress that for the universality proof above, we considered connections between the pairs $(Q_1,Q_2)$ and $(Q_1,Q_3)$: this was done merely for a sake of simplicity to reduce the proof the the Toffoli’s proof of Sec.~\ref{sec:universality}. In the Figs.~\ref{fig:alternative1}, \ref{fig:alternative2} this cannot be done by construction. Therefore the pairs employed are $(Q_1,Q_3)$ and $(Q_4,Q_8)$.
However the proof can be generalized to the case of any pair of qubits provided we connect two qubits of the same parity and two qubits of different parity (like the arrangement we used in the figures, i.e. the $(Q_1,Q_3)$ and $(Q_4,Q_8)$ pairs).

\FloatBarrier


\begin{thebibliography}{99}
\bibitem{projectors}
If the index  $i$ identifies a regular $\chi$-type qubit we have  
$\hat{P}_{\langle i \rangle}:= |g g \rangle\langle g g|$ and $\hat{Q}_{\langle i \rangle}:= |ee \rangle\langle ee| +
|eg \rangle\langle eg| + |ge \rangle\langle ge|$, where  $|g g \rangle$, $|e g \rangle$,
$|g e \rangle$, and $|ee \rangle$
represent the energy levels of the two qubits of $\overline{\chi}$-type that exhibit a $ZZ$ coupling with such site. 
On the contrary, if $i$ identifies a $B$-type ($A$-type) element with a black triangle then there are three interacting $B$($A$)-type sites so that $\hat{P}_{\langle i \rangle}:= |g g g\rangle\langle g g g|$ and $\hat{Q}_{\langle i \rangle}:= |eee \rangle\langle eee| +
|eeg \rangle\langle eeg| + \cdots + |gge \rangle\langle gge|$.
\end{thebibliography}
\end{document}